\documentclass{an_art}
\input{psfig.sty}
\begin{document}

\begin{titlepage}
\setcounter{page}{1}
\headnote{Astron.~Nachr. ... (...) }
\makeheadline

\title{Internal shocks in blazar jets}

\author{{\sc Gabriele Ghisellini}\\
Osservatorio Astronomico di Brera, Via Bianchi 46, I-23807 Merate Italy
}

\date{Received 1999 June 7; accepted ....} 
\maketitle

\summary
The discovery of strong $\gamma$--ray and hard X--ray emission 
and the results from various multifrequency campaigns have disclosed 
new aspects of the blazar phenomenology, leading to a much more
robust understanding of the mechanisms underlying their emission, 
and offering clues for the energetics of their relativistic jets.
I review these aspects, and propose that all the
diversity in the blazar phenomenology depends on only one parameter.
I also suggest that some of the blazar characteristics can be 
explained by the internal shock scenario, as proposed to
explain the emission from $\gamma$--ray bursts.
END
\keyw
blazars; radiation processes, jets, $\gamma$--rays.
END
\end{titlepage}

\section{Introduction}
Fossati et al. (1998) and Ghisellini et al. (1998) proposed that the
blazar diversity can be framed in a coherent picture, where blazars
form a sequence according to their observed bolometric luminosity.
All blazars have a spectral energy distribution (SED) with two peaks
in a $\nu$--$\nu F_\nu$ plot: we believe that the low frequency peak
(between the far IR and X--ray band) is due to synchrotron emission,
and the high energy peak (between the MeV and TeV bands) to inverse
Compton emission produced by the same electrons.  As the luminosity increases, both peaks shift to lower frequencies,
and the inverse Compton power increases its dominance (i.e. the ratio
between the inverse Compton $\gamma$--ray luminosity and the
synchrotron power, $L_c/L_s$).
By fitting the SED of blazars with a synchrotron--self Compton model
which includes the possible contribution of radiation produced outside
the jet (namely radiation directly from the broad line region, or
scattered by intercloud material, or from the walls of the jet), we
can estimate the physical quantities inside the emitting region, such
as the magnetic field, the emitting particle density and the shape of
their energy distribution, together with the Lorentz factor of the
bulk motion of the region and its dimension.
Even if the results may be affected by large uncertainties in single
sources, due mainly to the paucity of simultaneous spectra, they are
meaningful when derived for a relatively large sample of sources in a
statistical sense.
One of the most important parameters is the random Lorentz factor,
$\gamma_{peak}$, of the particles emitting most of the radiation,
i.e. the ones responsible for the emission at both peaks of the SED.
We find that $\gamma_{peak}$ strongly correlates with the intrinsic
energy density $U$ (both radiative and magnetic), the total intrinsic
power, and the amount of Compton dominance.
This both unifies and explains the different sub--classes of blazars:
low luminosity objects are characterized by small energy densities and
large values of $\gamma_{peak}$, producing a SED peaking at high
energies, as the one typical of ``high frequency peaked" BL Lac
objects (Giommi \& Padovani 1994).
External radiation (e.g. emission lines) is absent in these sources,
and this results into $L_c/L_s$ of order unity or less.
At the other extreme, powerful blazars with strong emission lines are
characterized by large values of $U$ and small $\gamma_{peak}$: their
SED peaks at low energies, and the presence of external radiation
boosts the inverse Compton process, producing highly Compton dominated
sources.


But where and how is this radiated energy produced?
It is widely believed that the radiation in the jet comes
from the transformation of kinetic energy of the bulk motion into
random energy.
Here I will suggest that several characteristic of blazars, including
those just described, can be explained by an {\it internal shock
scenario}, where shells of slightly different velocities interact at
some distance from the central engine.

\section{Internal shocks}
Rees (1978) faced the problem of explaining the optical emission of
the jet of M87. 
The large energies of the emitting electrons imply very short
cooling times, excluding the possibility of transporting those electrons
from the nucleus.
In situ acceleration is necessary.  
A way to achieve this is to envisage a central engine
working not continuously, but producing different shells of material
moving with different bulk Lorentz factors $\Gamma$.  
A faster shell catches up the
slower one at a distance of the order $R_i\sim \Gamma^2 R_0$, if the
slower moves with $\Gamma$, and the faster with $2\Gamma$ and 
$R_0$ is the initial separation of the two shells.  
Curiously, this idea was more or less forgotten in the blazar field
(but see Sikora, Begelman \& Rees 1994), 
but became the standard scenario to explain the emission of $\gamma$--ray 
bursts (GRB; Rees \& Meszaros 1994). 

Note that internal shocks are different from the so called {\it external
shock scenario} --  proposed to explain the afterglow emission of GRBs, and 
recently considered by Dermer (1999) to also explain the blazar 
emission -- where  the $\gamma$--ray fireball/'blob' in blazars 
{\it decelerates} interacting with the interstellar medium.
I will present in the following some arguments in favor of the
internal shock scenario operating in the blazar jet.

\subsection{Shock location}  

We know, mainly thanks to EGRET onboard CGRO, that most of the
emission of blazars is in the $\gamma$--ray band.
This emission, and the fact that it varies with short timescales,
implies that there must be a preferred location where dissipation
of the bulk motion energy occurs. On one hand, 
if it were at the base of the jet, and hence close to the accretion disk,
the produced $\gamma$--rays should be inevitably absorbed by 
photon--photon collisions, with associated copious pair production,
reprocessing the original power from the $\gamma$--ray to the X--ray
part of the spectrum (contrary to observations).
On the other, if it were far away, in a large region of the jet,
it becomes difficult to 
explain the observed fast variability,
even accounting for the time--shortening due to the Doppler effect.
Regions distant $\sim 10^{17}$ cm from the base of the jet are
then selected.
This distance is likely to be within the broad line region: as
the extra seed photons provided by emission lines enhance
the efficiency of the Compton process responsible for the
$\gamma$--ray emission. 
And indeed this is the typical distance at which 
two shells, initially 
separated by $R_0\sim 10^{15}$ cm
(comparable to a few Schwarzschild radii), moving with $\Gamma \sim 10$ 
would collide.

\subsection{Time variability}
The light curve of GRBs is extremely variable and
spiky, with the shortest spikes lasting for 1 ms.
Suppose that a single spike corresponds to a single shell--shell encounter,
and that its duration is determined by the time needed to
the two shells to cross each other. 
If the shell widths is of the same order of their separation, this
is given by the time elapsed to go from $R_i$ to $2 R_i$.
But the observer will see the crossing time Doppler contracted
by the factor $(1-\beta\cos\theta)\sim \Gamma^{-2}$
(where $\theta\sim 0^\circ$ is the viewing angle), since the shell
is moving almost at the speed of light.
The shortest spikes of the GRB light curves
can be therefore  associated to the size of the central engine (i.e. $R_0$),
since $t_{obs}\sim R_i/(c\Gamma^2) \sim R_0/c$.

If the mechanism powering GRB and blazar emission is the same,
we should expect a similar light curve from both systems, 
but with times appropriately
scaled by the different $R_0$, i.e. the different masses
of the involved black holes.
Indeed, if one stretches a GRB light curve by a factor $\sim 10^8$,
one obtains light curves very similar to the blazar ones,
with the shortest timescales of the order of hours--days.
In addition, by examining  GRB light curves through Fourier analysis and
adding up the resulting power spectra, a remarkably
well defined power law (in frequency space) is found, with slope 
$-5/3$, characterizing a Kolmogoroff spectrum (Belobodorov, Stern \& Svensson 1998).
This may be also the case for blazars.

\subsection{Efficiency}
Most of the power transported by the jet must reach the
radio lobes. Hence only a small fraction can be radiatively dissipated.
The efficiency $\eta$ of two blobs/shells for converting ordered into
random energy depends on their masses $m_1$, $m_2$ and 
bulk Lorentz factors $\Gamma_1$, $\Gamma_2$, as
\begin{equation}
\eta \,=\, 1-\Gamma_f\, { m_1+m_2\over \Gamma_1 m_1 +\Gamma_2m_2}
\end{equation}
where $\Gamma_f=(1-\beta_f^2)^{-1/2}$ is the bulk Lorentz factor 
after the interaction and is given by 
(see e.g. Lazzati, Ghisellini \& Celotti 1999)
\begin{equation}
\beta_f = {\beta_1\Gamma_1m_1+ \beta_2\Gamma_2m_2 \over
\Gamma_1m_1+ \Gamma_2m_2}
\end{equation}
The above relations imply, for shells of equal masses and 
$\Gamma_2=2\Gamma_1=20$, $\Gamma_f=14.15$ and $\eta=5.7\%$.

Efficiencies $\eta$ around 5--10\% are just what needed
for blazar jets.
Furthermore, we can ``directly" estimate the values appropriate for
blazars from the amounts of power radiated and transported 
by jets in the form of cold protons,
magnetic field and hot electrons and/or electron--positron pairs.
This requires to determine the bulk Lorentz factor,
the dimension of the emitting region, the value of the magnetic
field and the particle density. 
Ghisellini et al. (1998) have computed these
quantities for all (51) blazars
detected by EGRET for which both the redshift and the $\gamma$--ray
spectral shape are currently known, and thus determined 
%
\begin{equation}
L_p \, = \, \pi R^2 \Gamma^2\beta c\, n_p^\prime m_p c^2;\qquad
L_e\,  = \, \pi R^2 \Gamma^2\beta c\, n_e^\prime \langle \gamma \rangle 
m_e c^2 ;\qquad
L_B\, = \, \pi R^2 \Gamma^2\beta c \, {B^2 \over 8\pi} 
\end{equation}
These powers can be compared with the radiated one estimated in the
same frame (in which the emitting blob is seen moving). The power
radiated {\it in the entire solid angle} is thus $L=L^\prime\Gamma^2$.
All these quantities are plotted in Fig. 1 (Celotti \& Ghisellini 1999, 
in prep.), from which it can be noted that:
\begin{figure}
\psfig{file=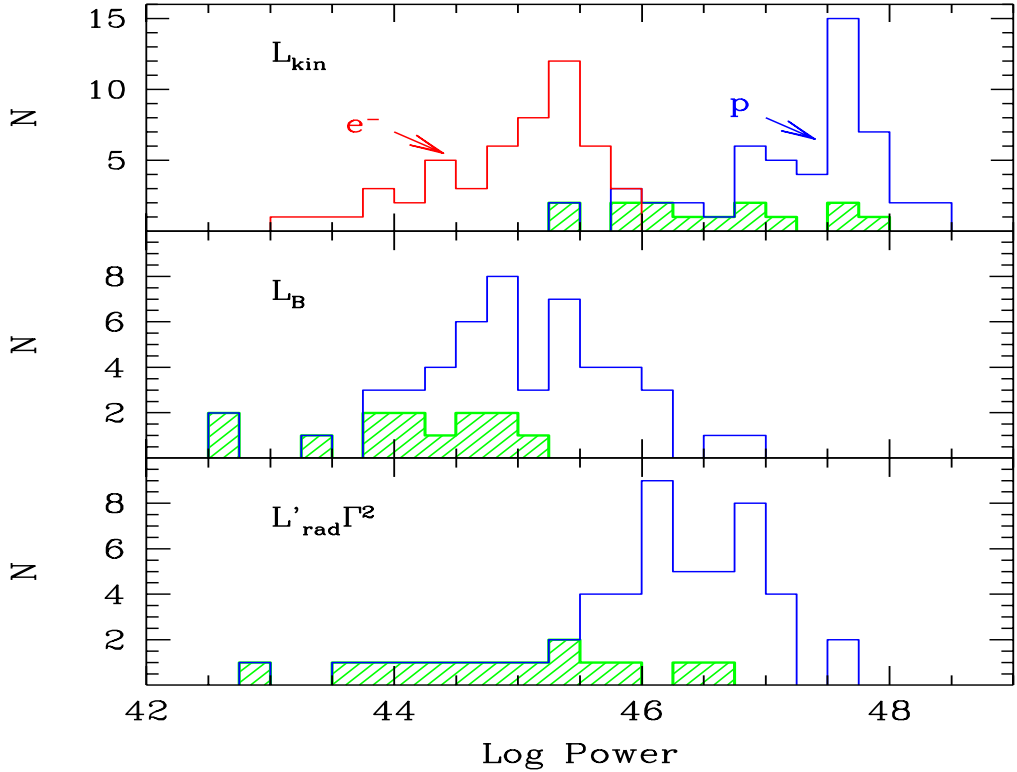,width=14truecm,height=11truecm}
\vskip -0.8 true cm
\caption{Histograms of the power transported in the form
of cold protons and relativistic electrons (upper panel),
Poynting vector (middle panel) and the emitted luminosity (bottom panel).
Shaded areas correspond to BL Lac objects in all histograms
but $L_e$ (in this case the shaded area has been omitted for clarity). From 
Celotti \& Ghisellini 1999, in prep.}
\end{figure}
\begin{itemize}
\item If the jet is formed by electrons--positrons only, the corresponding 
kinetic power is smaller than the radiated power,
posing a serious energy budget problem.
\item If there is a proton for each electron, the bulk kinetic
power is $\sim$10 times larger than the radiated one.
This value agrees with the energy requirements of the radio lobes.
\item The power in the Poynting flux is of the same order of $L_e$,
indicating that the magnetic field is close to the equipartition with
the electron energy density.
This suggests that the magnetic field is not a prime energy carrier, but
that is a sub--product of the interaction transforming
bulk into random energy.
\item The power in electrons is smaller than the radiated power,
by a factor $\sim$10.
This requires  reacceleration/continuous injection of electrons.
\end{itemize}

\subsection{Spectrum}
The internal shock scenario allows to estimate the total energy
available for radiative dissipation.
In fact, in the rest frame of the fast shell, the bulk kinetic energy
of each proton of the slower shell is $\sim (\Gamma^\prime-1)m_pc^2$,
where $\Gamma^\prime\sim 2$.  This is what can be transformed into
random energy.
Assume now that the electrons share some of this available energy
(through an unspecified acceleration mechanism).
In the comoving frame, the acceleration rate can be written as 
$\dot E_{heat} \sim (\Gamma^\prime-1)m_p c^2 /t^\prime_{heat}$.
The typical heating timescale may correspond to the time needed for the 
two shells to cross, i.e. 
$t^\prime_{heat}\sim \Delta R^\prime/c\sim R/(c\Gamma)$,
where $\Delta R^\prime$ is the shell width (measured in the same frame).
The heating and the radiative cooling rates will balance for some
value of the random electron Lorentz factor $\gamma_{peak}$:
\begin{equation}
\dot E_{heat}\, =\,  \dot E_{cool}\, \to \,
{\Gamma m_p c^3 \over R} \, =\, {4\over 3} \sigma_T c U\gamma^2_{peak}
\, \to \,
\gamma_{peak} \, =\, \left({ 3\Gamma m_p c^2 \over 4 \sigma_T R U}\right)^{1/2}
\end{equation}
This prediction can be tested against the parameters derived
by the fitting of the EGRET blazars, described above, and the result is
shown in Fig. 2. The agreement is surprisingly good both in the 
relative dependence and in the absolute normalization.

\section{Conclusions}
The idea that blazar emission can originate from internal shocks 
seems promising. 
In addition, the similarity with GRBs phenomenology 
suggests that internal shocks
might be one chief mechanism to convert relativistic bulk motion into
radiation.

\begin{figure}
\vskip -0.7 true cm
\psfig{file=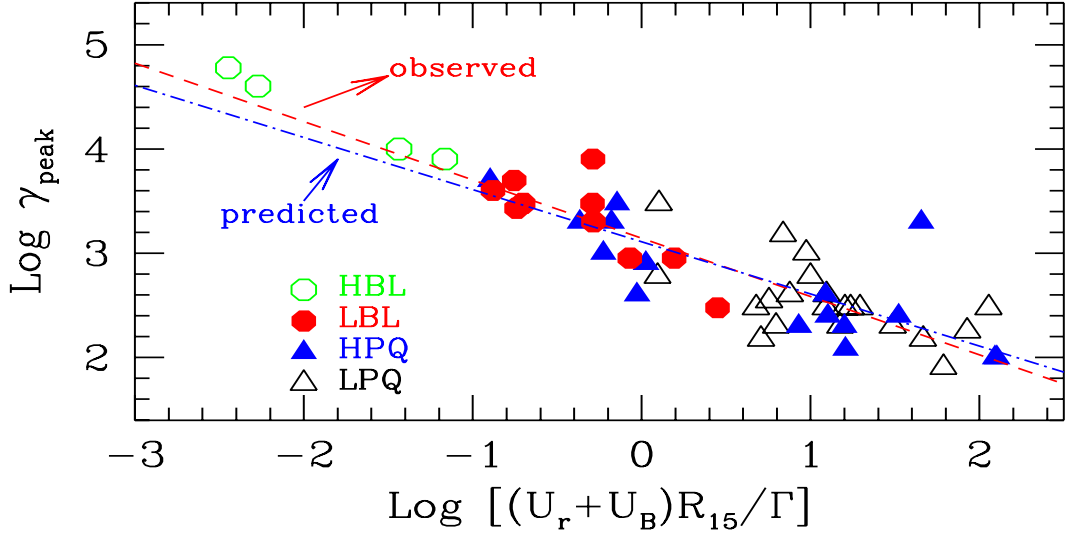,width=15truecm,height=13truecm}
\vskip -4.5 true cm
\caption{Correlation between the Lorentz factor of the electrons emitting
at the peaks of the SED (both synchrotron and inverse Compton) and
the quantity $U R/\Gamma$, where $U=U_r+U_B$ is the sum of radiation and
magnetic energy densities, $R/\Gamma$ is the shell width ($R=10^{15}R_{15}$ cm). 
Labels indicate 
High and Low frequency peak BL Lac objects (HBL and LBL),
and High and Low polarization quasars (HPQ and LPQ, respectively).  }
\end{figure}

\acknowledgements 
It is a pleasure to thank Annalisa Celotti, for the ongoing collaboration
within which the ideas reported in this work have been conceived.

\addresses
\rf{Gabriele Ghisellini, 
Osservatorio Astronomico di Brera, 
Via Bianchi 46, I-23807 Merate, 
Italy, e--mail: gabriele@merate.mi.astro.it}

\end{document}